\documentclass[%
 reprint,
 amsmath,amssymb,
 aps,
 pra,
floatfix,
]{revtex4-2}

\usepackage{graphicx}% Include figure files
\usepackage{dcolumn}% Align table columns on decimal point
\usepackage{bm}% bold math
\usepackage{amsmath}
\usepackage[colorlinks=true, citecolor=blue, linkcolor=blue, urlcolor=blue]{hyperref}
\begin{document}

\raggedbottom

\preprint{APS/123-QED}

\title{Optimal Control Design of Robust Raman Pulses for High-Fidelity Cold-Atom Interferometry}% Force line breaks with \\
% \thanks{...} 会为标题添加一个脚注符号(例如星号*)，并在页面底部显示脚注内容。
% 通常用于对标题的补充说明或一些特殊的致谢。

\author{Ziwen Song}
%  \email{15235225125@163.com}
\email{songzw24@mails.jlu.edu.cn}
%  \homepage{http://www.Second.institution.edu/~Charlie.Author}
 \affiliation{Jilin University, College of Instrumentation and Electrical Engineering, Changchun, People’s Republic of China}%Lines break automatically or can be forced with \\
%  \email{Second.Author@institution.edu}
% \affiliation{%
%  Authors' institution and/or address\\
%  This line break forced with \textbackslash\textbackslash
% }%

% \collaboration{MUSO Collaboration}%\noaffiliation

%  \homepage{http://www.Second.institution.edu/~Charlie.Author}
% \affiliation{
%  Second institution and/or address\\
%  This line break forced% with \\
% }%
% \affiliation{
%  Third institution, the second for Charlie Author
% \affiliation{%
%  Authors' institution and/or address\\
%  This line break forced with \textbackslash\textbackslash
% }%

% \collaboration{CLEO Collaboration}%\noaffiliation

\date{\today}% It is always \today, today,

\begin{abstract}
The performance of high-precision cold-atom interferometers is often limited by imperfections in the Raman laser fields. We present a reproducible framework for robust Raman mirror-pulse design and compare Krotov, GRAPE, and CRAB under a common normalized peak-amplitude limit of 3.0. The effective two-level model uses a 25-member detuning--amplitude ensemble and a dense out-of-sample grid. In the fixed-budget study, GRAPE attained a terminal ensemble error of $1.224\times10^{-2}$ and projected Krotov attained $1.243\times10^{-2}$; Krotov occupied a slightly larger $P_e\ge0.9$ grid fraction (0.177 versus 0.170). The best of five CRAB seeds gave $3.081\times10^{-2}$. In the interferometer calculation, GRAPE produced the largest contrast, $0.582\pm0.019$, while Krotov and the selected CRAB pulse gave $0.454\pm0.019$ and $0.439\pm0.023$, respectively. These results establish a reproducible comparison of trade-offs, rather than universal superiority of any one optimizer.
% \begin{description}
% Secondary publications and information retrieval purposes.
% You may use the \texttt{description} environment to structure your abstract;
% use the optional argument of the \verb+\item+ command to give the category of each item. 
% \end{description}
\end{abstract}

\maketitle

%\tableofcontents

\section{Introduction}

Precision gravity measurement is an indispensable core technology in frontier fields such as geodesy\cite{Flechtner_2021}, geophysical exploration\cite{Jensen_2025}, tests of fundamental physics\cite{Rosi_2014, Lamporesi_2008}, and inertial navigation\cite{Canuel_2006, Cheiney_2018, Dickerson_2013}. With the continuous advancement of science and technology, increasingly stringent requirements are being placed on the precision of measuring the gravitational field and its dynamic variations. Among the numerous measurement techniques, quantum gravimeters based on the principle of cold-atom interferometry utilize the coherent interaction between lasers and atomic wave packets to precisely map inertial information, such as gravitational acceleration, onto the quantum phase of the atoms\cite{Kasevich_1991}. Owing to their extremely high theoretical sensitivity, long-term stability, and measurement accuracy, they have become one of the most promising technical approaches in this domain, demonstrating significant advantages in areas such as determining the fundamental gravitational constant\cite{Ekstrom_1995, Bouchendira_2011} and testing the equivalence principle\cite{Bonnin_2015, Asenbaum_2020}.

A typical Mach-Zehnder (MZ) atom interferometer employs a sequence of laser pulses—commonly a $\pi/2$-$\pi$-$\pi/2$ sequence of Raman pulses—to coherently manipulate the atomic wave packet through splitting, reflection, and recombination. Under ideal conditions, the output atomic population exhibits a clear sinusoidal relationship with the phase difference accumulated along the interferometric paths, resulting in a fringe contrast approaching 100\%. However, in a realistic experimental environment, noise and imperfections in the laser system are key bottlenecks that limit the interferometer's actual performance\cite{Wu_2019}. Laser frequency drifts, intensity fluctuations, and errors in pulse duration all prevent the Raman pulses from precisely controlling the atomic states, thereby reducing the pulse fidelity. Such control errors directly degrade the coherence of the interferometric process, leading to a significant reduction in the final fringe contrast and severely affecting the signal-to-noise ratio and ultimate precision of the gravity measurement\cite{Szigeti_2012, Le_Gou_t_2008}. Therefore, designing laser pulses that are robust against experimental noise while maintaining high fidelity is a central challenge for further enhancing the performance of atom interferometers.

To address this challenge, pulse-shaping techniques originating from the field of nuclear magnetic resonance (NMR), along with quantum optimal control theory, provide powerful tools for designing highly robust atomic manipulation schemes\cite{J_ger_2014, Daems_2013, van_Frank_2014}. The core idea is that by precisely modulating the amplitude and phase of the laser pulses over time, the quantum state can be guided to actively compensate for undesired phases introduced by external noise during its evolution, thereby reaching the target state with extremely high fidelity even in the presence of errors.

Building on the ideas discussed above, this paper applies quantum optimal control to the design of Raman mirror pulses in cold-atom interferometers. The Krotov algorithm is widely used in the field of quantum control because it can provide monotonic improvement under suitable update conditions\cite{Reich_2012, Glaser_2015, Saywell_2023}. We first establish a two-level model for the Raman interaction and then compare Krotov, GRAPE, and CRAB under the same initial pulse, amplitude limits, and robustness ensemble. Finally, through numerical simulations, we evaluate the resulting pulses under laser detuning and intensity fluctuations and propagate the best candidates through a Monte Carlo atom-interferometer model. The emphasis is on a reproducible and evidence-limited comparison rather than on assuming that one optimizer is superior in all metrics.

\section{Theoretical Framework}
\label{sec:theory}

\subsection{Quantum State Evolution}
\label{sec:evolution}

\begin{figure}[!htbp]
% 将图片宽度设置为当前分栏宽度的90%
\includegraphics[width=0.6\columnwidth]{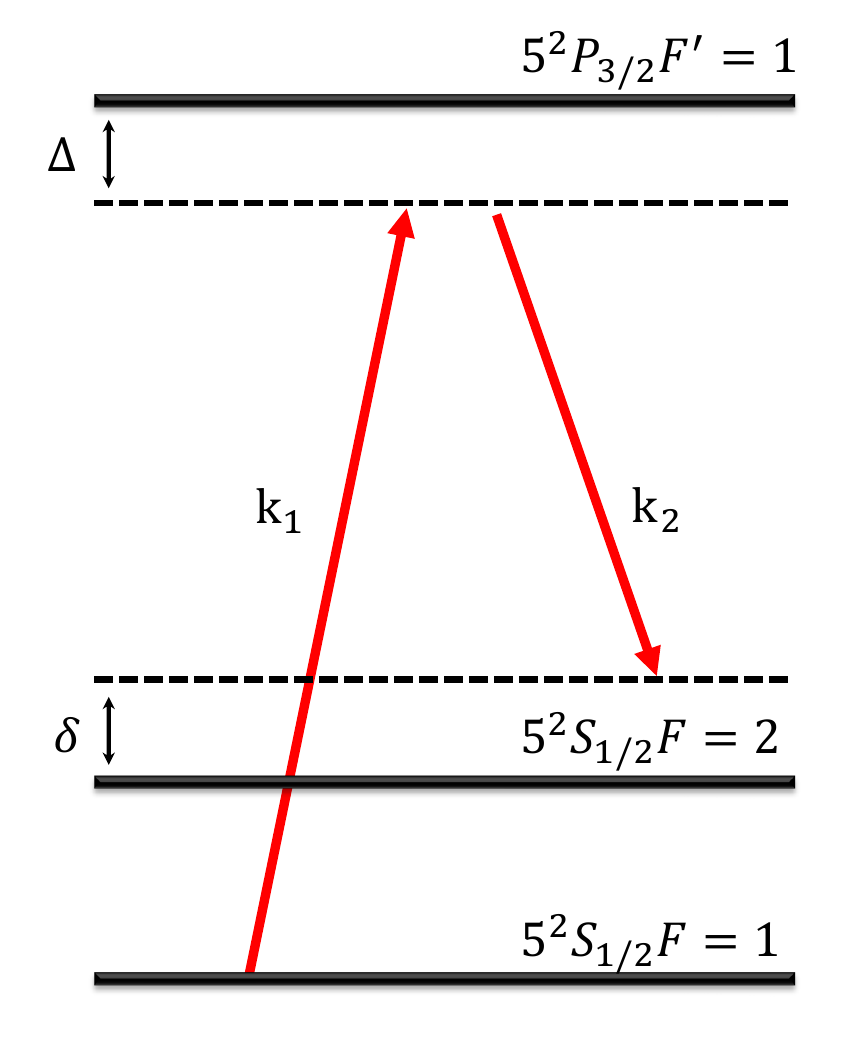}
\caption{\label{figure_1}
Energy level diagram for a two-photon stimulated Raman transition in an alkali atom (rubidium as an example). Two laser fields (with wave vectors $\mathbf{k}_1$ and $\mathbf{k}_2$) coherently couple the ground-state hyperfine levels. $\delta$ is the two-photon detuning, which corresponds to the detuning term in the effective two-level model.
}
\end{figure}

In gravimetry, atom interferometers are typically based on two-photon stimulated Raman transitions. We consider an alkali rubidium atom undergoing stimulated transitions between its hyperfine split energy levels. A stimulated Raman transition can be described as a two-photon process where two atomic states are coupled by two laser fields of different frequencies via an intermediate level. This process can be effectively approximated as a two-level system, as depicted in the energy level diagram in Fig.~\ref{figure_1}. We denote the ground state by $|g, \mathbf{p}\rangle$ (corresponding to $5^2S_{1/2}, F=1$) and the excited state by $|e, \mathbf{p}+\hbar\mathbf{k}_L\rangle$ (corresponding to $5^2S_{1/2}, F=2$). Using the Bloch sphere representation, this quantum state can be written as\cite{Shore_2011}:
\begin{equation}
    |\psi\rangle = \cos\left(\frac{v}{2}\right)|g\rangle + e^{i\phi}\sin\left(\frac{v}{2}\right)|e\rangle,
\end{equation}
where $v$ and $\phi$ are the polar and azimuthal angles on the Bloch sphere, respectively.

The Hamiltonian for this two-level atomic system interacting with the laser field is given by $H = \frac{\hbar}{2} \mathbf{\Omega}_R(t) \cdot \mathbf{\sigma}$, which can be expressed in matrix form as:
\begin{equation}
    H = \frac{\hbar}{2}
    \begin{pmatrix}
        \delta & \Omega_{\text{eff}} e^{-i\phi_L} \\
        \Omega_{\text{eff}} e^{i\phi_L} & -\delta
    \end{pmatrix}.
\end{equation}

The Schrödinger equation can then be expressed as:
\begin{equation}
    \begin{split}
        \frac{d}{dt}
        \begin{pmatrix} c_e(t) \\ c_g(t) \end{pmatrix}
        &= -i \frac{\hbar}{2} \mathbf{\Omega}_R(t) \cdot \mathbf{\sigma}
        \begin{pmatrix} c_e(t) \\ c_g(t) \end{pmatrix} \\
        &= -iH
        \begin{pmatrix} c_e(t) \\ c_g(t) \end{pmatrix}.
    \end{split}
\end{equation}
where $\mathbf{\sigma}$ represents the Pauli matrices and $\mathbf{\Omega}_R(t)$ is the Rabi vector, given by $\mathbf{\Omega}_R(t) = \Omega_{\text{eff}}(t)\cos(\phi_L)\mathbf{x} + \Omega_{\text{eff}}(t)\sin(\phi_L)\mathbf{y} + \delta(t)\mathbf{z}$. Here, $\Omega_{\text{eff}}(t)$ is the effective Rabi frequency and $\phi_L$ is the phase of the Raman laser. The magnitude of the Rabi vector, $\Omega(t) = \sqrt{\Omega_{\text{eff}}(t)^2 + \delta(t)^2}$, also depends on the two-photon detuning $\delta(t)$, which is given by\cite{Paul_1997}:
\begin{equation}
    \begin{split}
        \delta(t) ={}& (\omega_1(t) - \omega_2(t)) \\
                    & - \left(\omega_{eg} + \frac{\mathbf{p} \cdot \mathbf{k}_{\text{eff}}}{m} + \frac{\hbar |\mathbf{k}_{\text{eff}}|^2}{2m} + \delta^{ac}\right).
    \end{split}
\end{equation}

\begin{figure}[!htbp]
% 将图片宽度设置为当前分栏宽度的90%
\includegraphics[width=1\columnwidth]{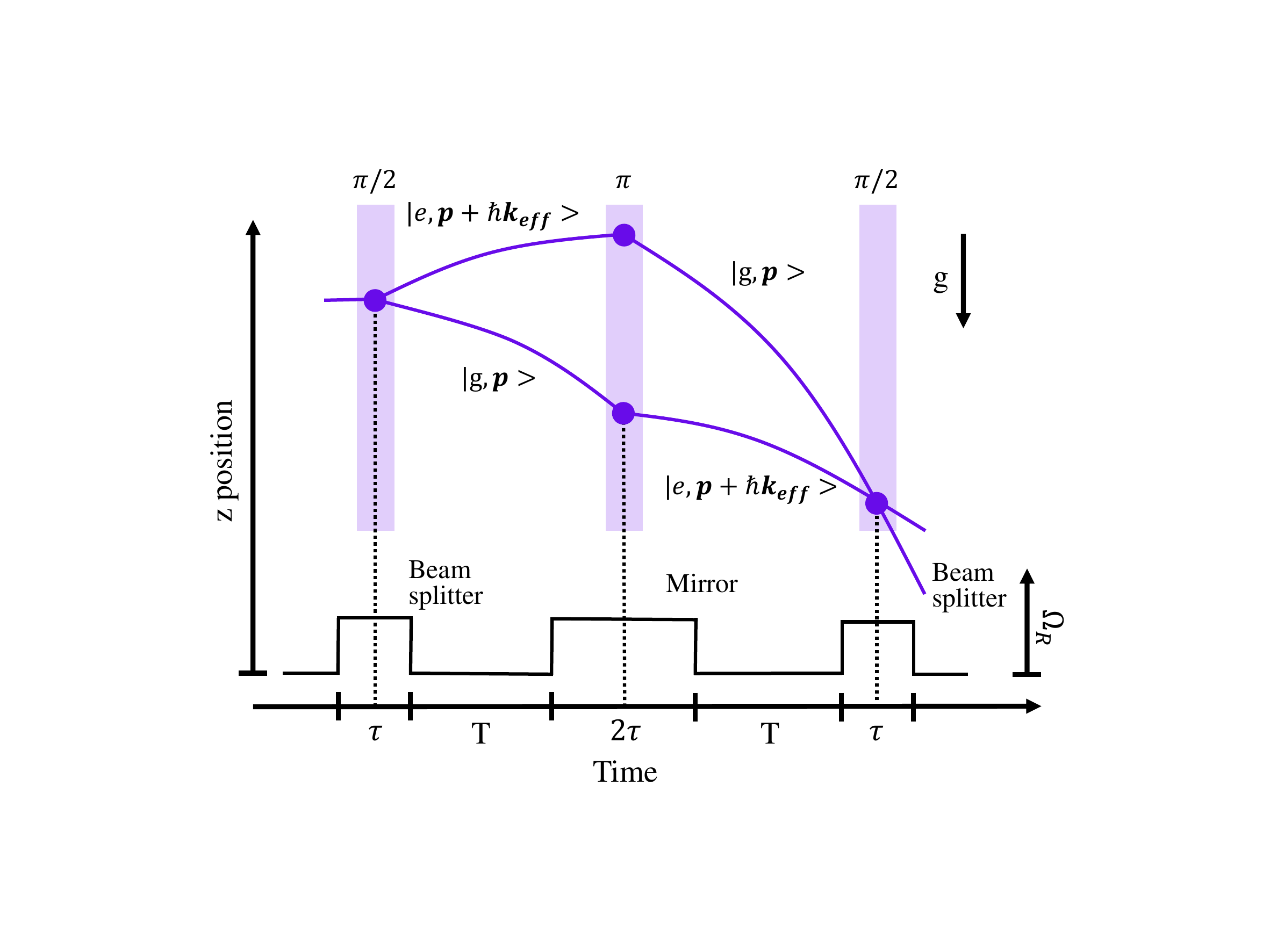}
\caption{\label{figure_2}
Spacetime path diagram for a Mach-Zehnder atom interferometer. The interferometer is formed by a $\pi/2$-$\pi$-$\pi/2$ sequence of Raman pulses, with a time interval of $T$ between pulses. The purple solid lines represent the two classical paths of the atomic wave packet under the influence of the gravitational field (g). Here, $|g, \mathbf{p}\rangle$ denotes the ground-state atom and $|e, \mathbf{p}+\hbar\mathbf{k}_L\rangle$ denotes the excited-state atom. The black curve below shows the timing of the Raman pulses, and $\Omega_R$ is the Rabi frequency.
}
\end{figure}

In this expression, $\mathbf{p} \cdot \mathbf{k}_{\text{eff}}/m = \mathbf{k}_{\text{eff}} \cdot \mathbf{v}$ is the Doppler shift, which contains information about the gravitational acceleration experienced by the atom cloud; $\hbar |\mathbf{k}_{\text{eff}}|^2 / 2m$ is the two-photon recoil shift, and $\delta^{ac}$ is the AC Stark shift. Figure~\ref{figure_2} shows the basic working principle of a Mach-Zehnder atom interferometer. Here, the atomic wave packet (purple solid lines) evolves in the gravitational field (g) and is coherently manipulated by a $\pi/2$-$\pi$-$\pi/2$ Raman pulse sequence (timing diagram below) for splitting, reflection, and recombination, ultimately forming a closed interferometric loop.

While the Schrödinger equation can be solved numerically, the quantum state evolution after applying a Raman pulse of duration $\Delta t$ can also be described by the propagator\cite{Saywell_2018}:
\begin{equation}
    U = e^{-iH \Delta t / \hbar} = 
    \begin{pmatrix} 
        C & -iS \\ 
        -iS^* & C^* \end{pmatrix},
\end{equation}
where C and S are defined as\cite{Dunning_2014}:
\begin{align}
    C ={}& \cos\left(\frac{\sqrt{\Omega_{\text{eff}}^2 + \delta^2}}{2} \Delta t\right) \nonumber \\ % <-- 在这里使用 \\ 进行换行
         & + i \frac{\delta}{\sqrt{\Omega_{\text{eff}}^2 + \delta^2}} \sin\left(\frac{\sqrt{\Omega_{\text{eff}}^2 + \delta^2}}{2} \Delta t\right), \\ % <-- 使用 & 对齐 + 号
    S ={}& e^{i\phi_L} \frac{\Omega_{\text{eff}}}{\sqrt{\Omega_{\text{eff}}^2 + \delta^2}} \sin\left(\frac{\sqrt{\Omega_{\text{eff}}^2 + \delta^2}}{2} \Delta t\right).
\end{align}

The transition probability $P_e$ after the application of the Raman pulse can be calculated from $P_e = |\langle e | U | \psi_0 \rangle|^2$\cite{L_pez_Monjaraz_2024}. For an arbitrary pulse shape, the analytical solution is complex. However, by employing the Magnus expansion to obtain a perturbative expansion in $\Delta t$, one can calculate an effective propagator for a given pulse's action on the two-level system\cite{Magnus_1954, Miao_2000}.
\begin{equation}
    \begin{split}
      U &= \exp(-i(H_0 + H_1 + H_2 + \dots)\Delta t / \hbar) \\
        &= U_0 U_1 \dots U_n \dots U_{N-1} U_N.
    \end{split}
\end{equation}
The propagator for each time step is treated as the action of a short rectangular pulse with a fixed laser amplitude and phase, acting on the atoms for a duration $dt$.

Therefore, the quantum state after evolving under the modulated pulse can be expressed as:
\begin{equation}
  |\psi_e\rangle = U |\psi_0\rangle.
\end{equation}

\subsection{Atomic Velocity Distribution}
\label{sec:velocity}

In an ideal atom interferometer, the Raman lasers have identical transverse intensity profiles, the laser frequency is stable, and the atomic cloud has a zero-velocity spread. Under these conditions, the maximum fringe contrast is obtained. When the velocity spread of the atom cloud is broader than the Raman transition linewidth, parasitic interferometric paths can emerge. These paths, which arise from atoms imperfectly following the intended trajectory shown in Fig.~\ref{figure_1} during the splitting and recombination processes, are a primary cause of reduced fringe contrast. Other contributing factors include imbalanced population splitting due to imperfect beam-splitter pulses and dephasing effects arising from velocity-dependent phase accumulation\cite{Saywell_2023, Luo_2016}.

In this section, we focus on the impact of the atomic velocity distribution on fringe contrast. In experiments, the Raman laser frequency is typically chirped to compensate for the acceleration of the atomic cloud's center-of-mass along the z-axis due to gravity. However, this chirp does not compensate for the relative motion between atoms arising from their thermal velocity spread. Therefore, we primarily consider the dephasing effects caused by this thermal motion. For a cold atom cloud undergoing thermal expansion, both the atomic positions and velocities can be modeled by Gaussian distributions. Assuming the position and velocity distributions are independent, the initial probability distribution for an atom with phase-space coordinates $(x_0, y_0, z_0, v_{x0}, v_{y0}, v_{z0})$ is given by\cite{Brzozowski_2002}:
\begin{equation}
    \mathcal{N}(x_0, y_0, z_0, v_{x0}, v_{y0}, v_{z0}) = \prod_{i \in \{x,y,z\}} g(i_0, \sigma_i) g(v_{i0}, \sigma_v).
\end{equation}

Here, $g(x,\sigma)=(2\pi\sigma^2)^{-1/2} \exp\left(-x^2 / (2\sigma^2)\right)$ is the general expression for a normalized one-dimensional Gaussian distribution, where $x$ corresponds to an initial position $i_0$ or velocity $v_{i0}$.

The temperature of the atom cloud affects the initial velocity distribution, leading to a velocity-dependent detuning $\delta(t)$. To account for this, one must integrate over the entire thermal velocity range to obtain the total transition probability\cite{L_pez_Monjaraz_2024, Paul_1997, Luo_2016}:
\begin{equation}
    P_{e} = \int |c_e(\mathbf{v})|^2 g(\mathbf{v}) d^3\mathbf{v},
\end{equation}
where $c_e(\mathbf{v})$ is the final probability amplitude of the excited state for an atom with initial velocity $\mathbf{v}$, and $g(\mathbf{v})$ is the initial atomic velocity distribution.

\subsection{Fringe Contrast}
\label{sec:contrast}

In a three-pulse Mach-Zehnder interferometer sequence, inertial effects such as gravitational acceleration induce a relative phase shift between the two paths of the interferometer. A final $\pi/2$ pulse converts this phase difference into a population difference between the final atomic states. For a counter-propagating beam configuration and in the limit of small detuning, the total transition probability is given by\cite{Stoner_2011, L_pez_Monjaraz_2024, Butts_2011}:
\begin{equation}
\begin{split}
    P_e ={}& |S_1|^2|S_2|^2|S_3|^2 + |C_1|^2|S_2|^2|C_3|^2 \\
    &+ |S_1|^2|C_2|^2|C_3|^2 + |C_1|^2|C_2|^2|S_3|^2\\
    & - 2\text{Re}\left[e^{i\Phi_g}C_1S_1^*(S_2^*)^2C_3^*S_3\right],
\end{split}
\end{equation}
where C and S are defined as above, and $\Phi_g$ is the relative phase induced in the interferometer. This expression can be simplified to the general form:
\begin{equation}
    P_e = \frac{1}{2}\left(A - B\cos(\Phi_g + \phi(\delta))\right).
\end{equation}
Here, $A$ represents the interferometer offset, and $\phi(\delta)$ is the relative phase shift introduced by the Raman pulses, given by $\phi S_{\pi/2}^1 + \phi S_{\pi/2}^3 - 2\phi S_\pi$. When the detuning $\delta$ is zero and no other phase noise is introduced, the expression simplifies to:
\begin{equation}
    P_e = \frac{1}{2}\left(A - B\cos(\alpha T^2)\right).
\end{equation}

To calculate the output transition probability of the interferometer, we use the propagator method. The total operator for the Mach-Zehnder interferometer sequence can be written as:
\begin{equation}
    U_{\text{Total}} = U_{\pi/2} U_{T1} U_{\pi} U_{T2} U_{\pi/2}.
\end{equation}
The total evolution can then be expressed as:
\begin{equation}
    |\psi_e\rangle = U_{\text{Total}} |\psi_0\rangle.
\end{equation}
The final output transition probability is calculated directly from the probability amplitude:
\begin{equation}
    P_e = |\langle e | U_{\text{Total}} | \psi_0 \rangle|^2.
\end{equation}
In the simulations below, the fringe contrast is calculated from a complete
scan of the final Raman phase as
\begin{equation}
    C = P_{\max}-P_{\min}.
\end{equation}
This definition is the peak-to-valley population modulation used by the
production code; it avoids substituting a single output population for a
fringe contrast.

This formalism, by directly calculating the final state via the time-evolution operator $U_{\text{Total}}$, avoids the reliance on the symmetry of the pulse sequence inherent in traditional methods. Consequently, it provides highly reliable predictions of the transition probability even under conditions involving asymmetric or complex pulses.

\subsection{Pulse-Shaping Techniques}
\label{sec:shaping}

Pulse-shaping techniques were first applied in the field of NMR. The core idea is to modulate the original pulse waveform by replacing the traditionally fixed amplitude and phase with time-varying functions. These functions are then designed using optimization methods to enhance performance metrics such as excitation efficiency, selectivity, and robustness. Amplitude modulation of the pulse alters its response to detuning, while phase modulation affects its sensitivity to errors in Rabi frequency and pulse duration. With modulation, the off-diagonal driving term of the Rabi vector can be expressed in a time-varying form as $\Omega_{\text{eff}}(t)[\cos\phi_L(t)\mathbf{x} + \sin\phi_L(t)\mathbf{y}]$, where $\Omega_{\text{eff}}(t)$ is the amplitude modulation and $\phi_L(t)$ is the phase modulation\cite{Wang_2024}.

Based on the description above, the system Hamiltonian becomes:
\begin{equation}
    H(t) = \frac{\hbar}{2} \left( \delta\sigma_z + \Omega_{\text{eff}}(t)[\cos\phi_L(t)\sigma_x + \sin\phi_L(t)\sigma_y] \right).
\end{equation}
In the Hamiltonian, $\delta$ represents the total detuning of the system, while the second term is determined by the control field, i.e., the instantaneous amplitude $\Omega_{\text{eff}}(t)$ and phase $\phi_L(t)$ of the pulse. Therefore, the physical essence of pulse-shaping techniques is to dynamically manipulate the direction and strength of the effective control field on the Bloch sphere by carefully designing the control functions $\Omega_{\text{eff}}(t)$ and $\phi_L(t)$. This steers the quantum state vector along a specific trajectory from an initial point to a target point. A robust evolution trajectory should be designed such that its endpoint is insensitive to variations in the uncontrollable components of the field, such as fluctuations in the detuning $\delta$ or an overall scaling of the control amplitude $\Omega_{\text{eff}}(t)$.

\section{Numerical Methods}
\label{sec:methods}

\subsection{Robust control objective}
\label{sec:optimization}

We expressed the two quadratures of the Raman field as independent real controls,
$\epsilon_x(t)=\Omega_{\mathrm{eff}}(t)\cos\phi_L(t)$ and
$\epsilon_y(t)=\Omega_{\mathrm{eff}}(t)\sin\phi_L(t)$.  A Gaussian pulse,
normalized such that $\int_0^{T_p}\Omega_{\mathrm{eff}}(t)\,dt=\pi$, provided
the common initial guess.  The reference angular frequency was
$\Omega_{\mathrm{ref}}=2\pi\times25$ kHz and the pulse duration was
$T_p=32~\mu$s.  Public parameters and exported controls were stored in SI
units.  Numerical optimization used
$\tilde t=\Omega_{\mathrm{ref}}t$,
$\tilde\Omega=\Omega/\Omega_{\mathrm{ref}}$, and
$\tilde\delta=\delta/\Omega_{\mathrm{ref}}$.

Robustness was included in the objective rather than assessed only after
optimization.  The training ensemble comprised the Cartesian product
\begin{align}
 \mu &\in\{0.90,0.95,1.00,1.05,1.10\},\\
 \tilde\delta &\in\{-0.20,-0.10,0,0.10,0.20\}.
\end{align}
where $\mu$ scales both control quadratures.  For each ensemble member $n$, the
terminal error was
\begin{equation}
 J_T^{(n)}=1-\left|\langle e|\psi_n(T_p)\rangle\right|^2,
\end{equation}
and the optimized objective was the unweighted ensemble average
$\overline{J}_T=N^{-1}\sum_n J_T^{(n)}$.

\subsection{Krotov optimization and comparison methods}

The Krotov update was evaluated with the state-to-state functional and an exact
two-level propagator.  The two controls shared the update envelope
\begin{equation}
 S(t)=\sin^2(\pi t/T_p),
\end{equation}
which suppresses updates at both pulse boundaries.  This replaces the Blackman
window stated in an earlier draft, which was not the window used by the
verified implementation.  The update penalties were
$\lambda=0.1$ with 100 projected updates under the common peak-amplitude limit. The field update followed
\begin{equation}
 \Delta\epsilon_j^{(k)}(t)=
 \frac{S(t)}{\lambda_j}
 \operatorname{Im}\left[
 \langle\chi^{(k-1)}(t)|
 \frac{\partial H}{\partial\epsilon_j}
 |\psi^{(k)}(t)\rangle
 \right],
\end{equation}
with forward states evaluated using the new field and co-states propagated
backward using the preceding iteration\cite{Palao_2003,Reich_2012}.

We compared projected Krotov with GRAPE and CRAB. GRAPE used the same piecewise-constant controls and ensemble objective, exact Fr\'echet derivatives, and a radial amplitude constraint\cite{Khaneja_2005}. CRAB parameterized corrections to the common Gaussian guess in a truncated randomized Fourier basis and used Nelder--Mead optimization\cite{Caneva_2011}. Five predetermined CRAB seeds were retained and summarized, while the best seed was used only for main-panel visualization. We report terminal error, wall-clock time, pulse total variation, root-mean-square spectral bandwidth, and the fraction of the evaluation grid above specified fidelity thresholds. Rectangular, Gaussian, and
super-Gaussian $\pi$ pulses, each independently area-normalized, served as
nonoptimized baselines.

\subsection{Interferometer simulation}

The Mach--Zehnder calculation applied an area-normalized
$\pi/2$--$\pi$--$\pi/2$ sequence to a Monte Carlo ensemble of
$^{87}$Rb atoms.  The mirror pulse was replaced by each candidate optimized or
standard pulse.  Thermal Doppler shifts, a fixed systematic two-photon
detuning, and transverse Gaussian intensity inhomogeneity were sampled
independently.  The third-pulse phase was scanned over $[0,2\pi]$, and the
fringe contrast was calculated from the complete scan as
\begin{equation}
 C=P_{\max}-P_{\min}.
\end{equation}
This procedure avoids the incorrect identification of a single output
population with fringe contrast.  Five fixed Monte Carlo seeds were used for
the production calculation, with 1000 atoms per seed.  Curves are reported as
the replicate mean with 95\% confidence intervals.

\subsection{Three-level validation and reproducibility}

The principal calculations used the effective two-level model.  We tested its
range of validity with a three-level lambda model for representative
$^{87}$Rb D2 parameters, using
$\Gamma=2\pi\times6.065$ MHz and single-photon detunings from 0.5 to 5 GHz.
The single-photon Rabi frequencies were chosen to reproduce
$\Omega_{\mathrm{ref}}$ in the large-detuning limit.  Transfer probability,
peak intermediate-state population, and effective-model error were recorded
in the Supplemental Material.  These parameters are representative values,
not a characterization of a specific apparatus.

All simulations were driven by one versioned configuration and used the
QuTiP and Krotov Python libraries\cite{Johansson_2012}.  Each output
contains the configuration hash, software versions, random seeds, controls,
source data, and stage-completion metadata.  Unit tests checked Hermiticity,
unitarity, probability conservation, analytic $\pi$- and $\pi/2$-pulse
limits, GRAPE gradients, Krotov monotonicity, CRAB reproducibility, and
interferometer fringe construction.

\section{Optimization Process and Pulse Performance Evaluation}
\label{sec:results}

\subsection{Optimization Performance (Analysis of Algorithm Convergence)}
\label{sec:convergence}

\begin{figure}[!htbp]
% 将图片宽度设置为当前分栏宽度的90%
\includegraphics[width=\columnwidth]{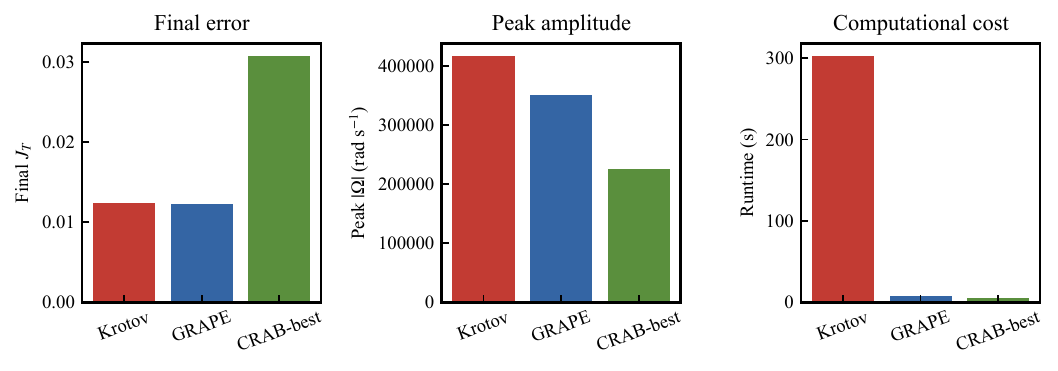}
\caption{\label{figure_3}
Final ensemble error and peak control resource for the constrained Krotov, GRAPE, and CRAB comparison. The projected Krotov value is the independently recomputed cost of the returned constrained control.
}
\end{figure}

We used the area-normalized Gaussian pulse as the common initial guess for all optimized calculations. The constrained production calculation used projected Krotov with $\lambda=0.1$, the same 25-member robustness ensemble for every method, and a common peak-amplitude limit of $3\Omega_{\rm ref}$. Figure~\ref{figure_3} reports final constrained-control performance rather than a formal monotonic-convergence trace.

The final ensemble errors were $1.243\times10^{-2}$ for projected Krotov and $1.224\times10^{-2}$ for GRAPE; neither reached the prescribed $10^{-3}$ target within this protocol. The best of five CRAB seeds gave $3.081\times10^{-2}$. The constrained comparison therefore provides evidence of trade-offs, not a universal optimizer ranking.

\begin{figure*}
% 将图片宽度设置为当前分栏宽度的90%
\includegraphics[width=\textwidth]{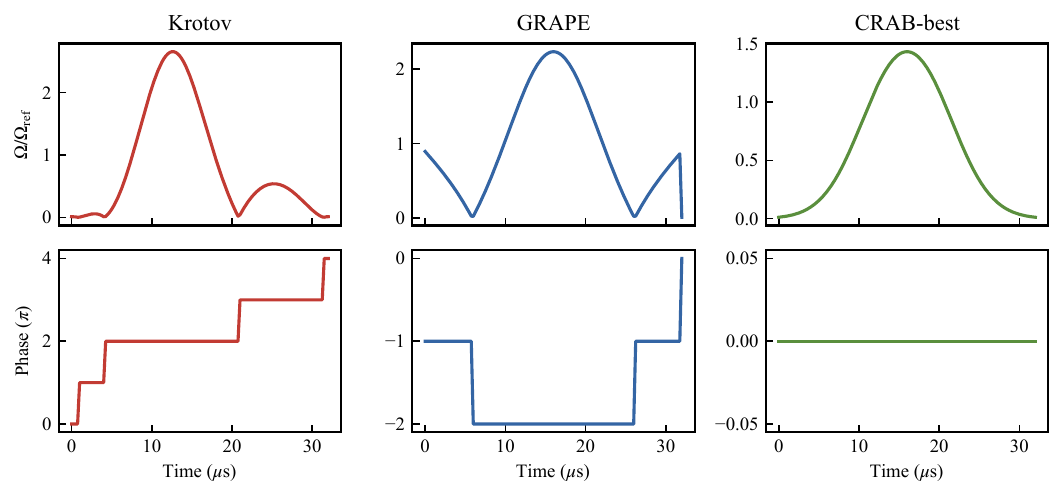}
\caption{\label{figure_4}
Amplitude and phase profiles for projected Krotov, GRAPE, and the selected best-of-five CRAB pulse. Complete Krotov waveform and spectrum are provided in the Supplemental Material.
}
\end{figure*}

All optimized controls use amplitude and phase modulation. They are shown symmetrically because they realize different trade-offs between terminal error, robustness, and control resource.

\subsection{Pulse Performance}
\label{sec:performance}

\begin{figure}[!htbp]
\includegraphics[width=0.8\columnwidth]{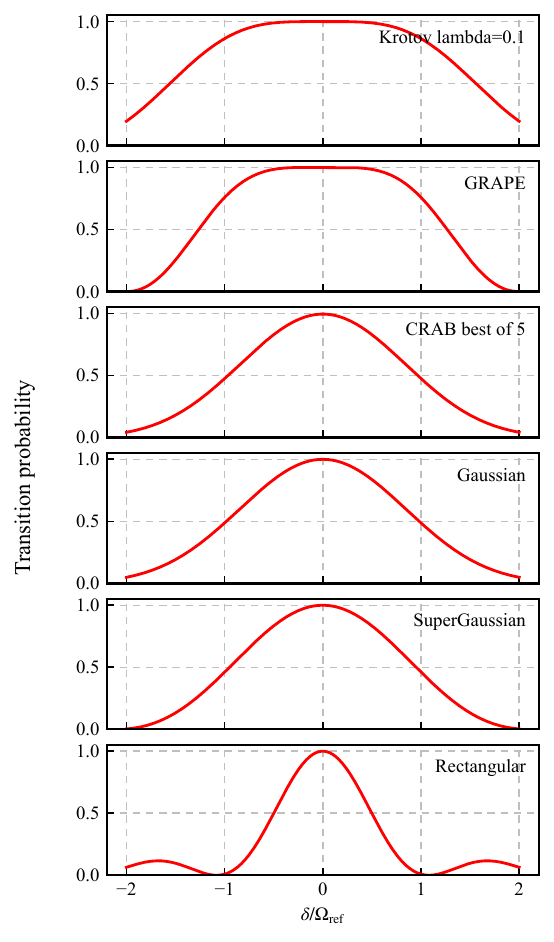}
\caption{\label{figure_5}
One-dimensional detuning response of projected Krotov, GRAPE, the selected best-of-five CRAB pulse, and three area-normalized standard pulses at $\mu=1$.
}
\end{figure}

We next evaluated each candidate on a dense grid that was not used for optimization. Figure~\ref{figure_5} shows the nominal-amplitude detuning response for all three optimized controls and the standard baselines.

The standard pulses retain high fidelity only near zero detuning. The optimized pulses trade waveform complexity for a flatter response over a wider interval, and their relative ranking depends on the metric and resource constraint.

\begin{figure}[!htbp]
\includegraphics[width=\columnwidth]{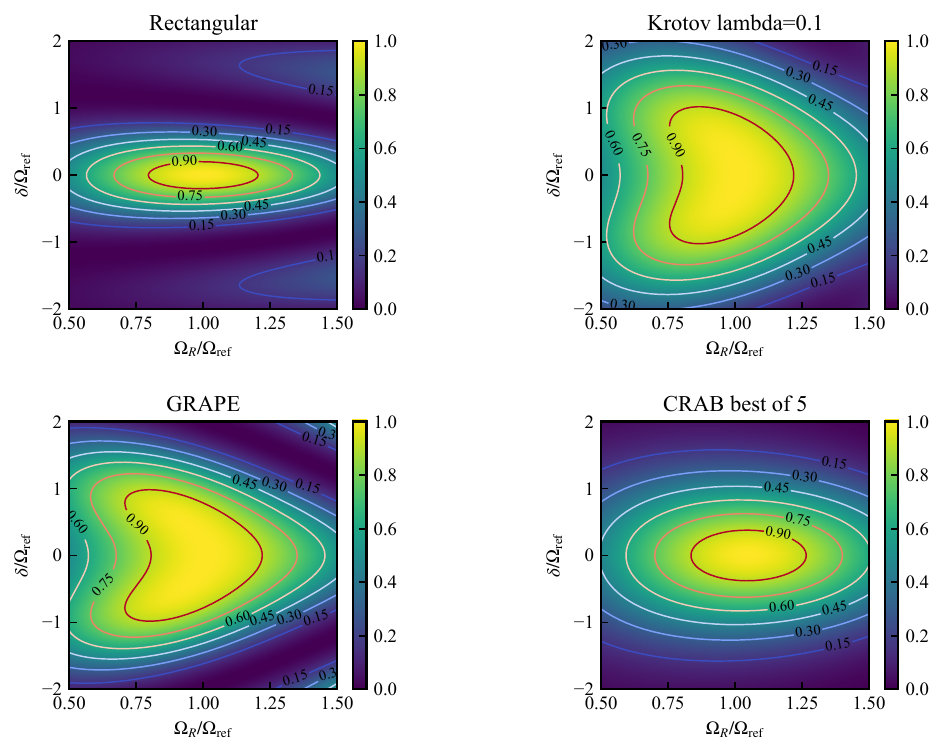}
\caption{\label{figure_6}
Two-dimensional robustness maps in detuning and amplitude-scale error.  The horizontal axis is $\Omega_R/\Omega_{\rm ref}$ and the vertical axis is $\delta/\Omega_{\rm ref}$.  Color and contours show the transition probability.  The panels compare rectangular, Gaussian, super-Gaussian, Krotov, GRAPE, and CRAB mirror pulses on the same dense grid.
}
\end{figure}

The two-dimensional maps in Fig.~\ref{figure_6} quantify simultaneous detuning and amplitude-scale robustness. The $P_e\ge0.9$ fractions are 0.177 for projected Krotov, 0.170 for GRAPE, 0.063 for the selected CRAB pulse, and 0.032 for the rectangular pulse. Thus the common peak-amplitude limit changes the algorithmic ranking and motivates a multi-metric comparison.

\begin{figure}[!htbp]
\includegraphics[width=1\columnwidth]{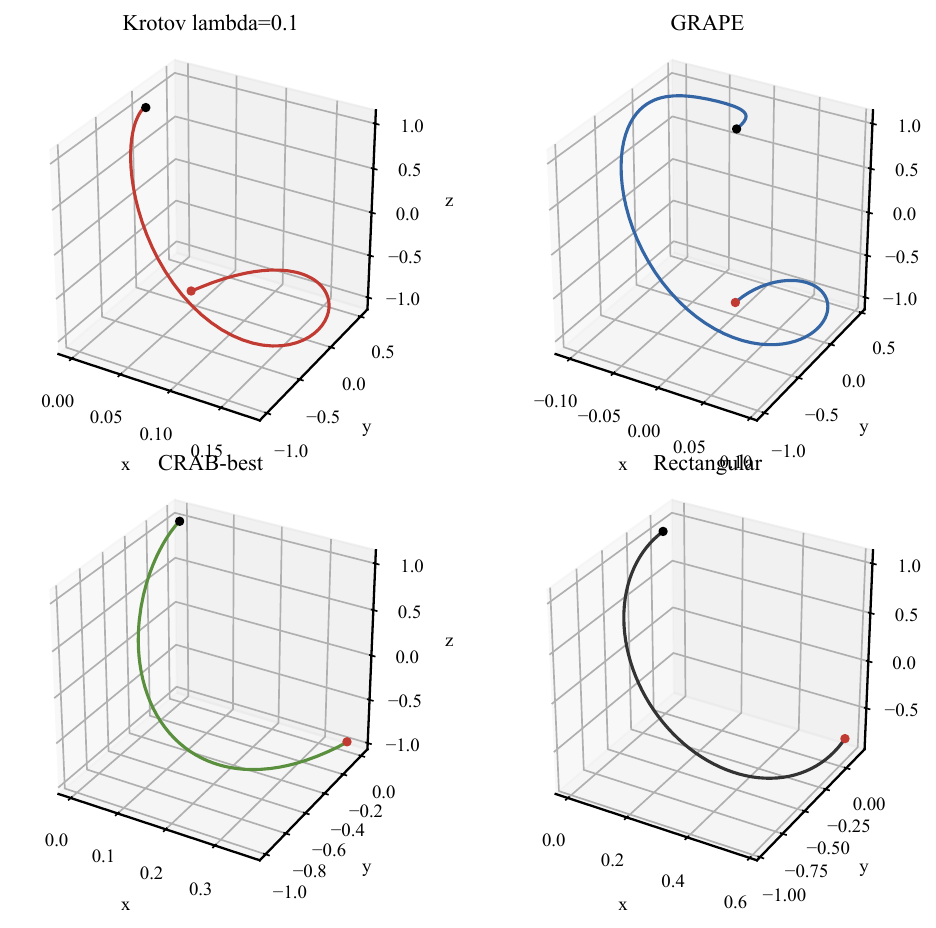}
\caption{\label{figure_7}
Bloch-vector trajectories for projected Krotov, GRAPE, the selected CRAB pulse, and a rectangular pulse at the common detuning $\delta/\Omega_{\rm ref}=0.2$.
}
\end{figure}

To illustrate the control mechanisms, we simulated all optimized trajectories and the rectangular reference at the same detuning, as shown in Fig.~\ref{figure_7}. The shaped controls follow distinct non-planar paths, whereas the rectangular pulse follows a detuning-tilted single-axis rotation.

\subsection{Gravimetry Performance}
\label{sec:gravimetry}

Ultimately, the performance of the Raman pulses must be tested by their practical application in an atom interferometer. In a Mach-Zehnder atom interferometer, the fidelity and robustness of the beam-splitter ($\pi/2$) and mirror ($\pi$) pulses directly determine the contrast of the final interference fringes. A higher contrast implies a stronger signal and a higher signal-to-noise ratio, which in turn enables more sensitive phase measurements.

\begin{figure}[!htbp]
\includegraphics[width=0.9\columnwidth]{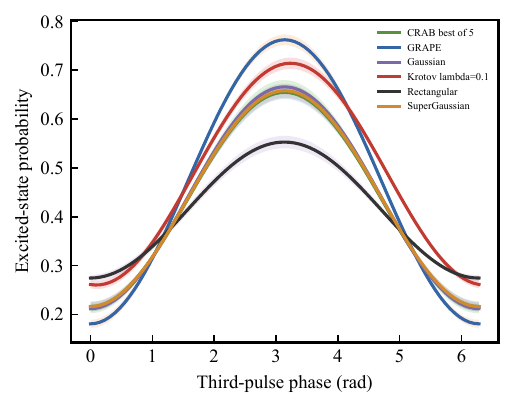}
\caption{\label{figure_8}
Simulated interferometer fringes under thermal Doppler broadening, transverse Gaussian intensity variation, and a fixed systematic detuning $\delta/\Omega_{\rm ref}=0.2$.  The mirror pulse is varied while the two beam splitters are area-normalized $\pi/2$ pulses.  Curves show the Monte Carlo mean over five fixed seeds, and shaded regions show 95\% confidence intervals.
}
\end{figure}

To link pulse robustness to an interferometric observable, we simulated the full $\pi/2$--$\pi$--$\pi/2$ sequence with each candidate as the mirror pulse. The fringe contrast was calculated from the complete phase scan as $C=P_{\max}-P_{\min}$. The resulting contrasts are $0.454\pm0.019$ for projected Krotov, $0.582\pm0.019$ for GRAPE, and $0.439\pm0.023$ for the selected CRAB pulse. The rectangular, Gaussian, and super-Gaussian baselines give $0.279\pm0.017$, $0.454\pm0.023$, and $0.442\pm0.024$, respectively. Under this constrained protocol, GRAPE gives the largest simulated contrast.

This result connects single-pulse robustness to an interferometric observable.
The magnitude and uncertainty of the contrast improvement are reported directly
from the Monte Carlo replicates rather than inferred from a single trajectory.

\section{Discussion}
\label{sec:discussion}

The comparison separates two questions that are often conflated in pulse
optimization: whether an algorithm reaches a low ensemble-averaged error and
whether the resulting field remains useful outside the discrete training
points.  Krotov, GRAPE, and CRAB employ different update rules and
parameterizations.  A comparison based only on iteration count would therefore
be misleading.  We instead report propagator evaluations, wall-clock time,
worst-case fidelity, spectral complexity, and performance on a common dense
grid.  These metrics expose trade-offs between numerical efficiency,
robustness, and experimental waveform complexity.

The effective two-level description is appropriate only when the single-photon
detuning is sufficiently large compared with the optical couplings and
linewidth.  Our three-level calculation quantifies this approximation for
representative parameters, but it does not model a particular instrument.
Spontaneous emission, differential AC Stark shifts, laser phase noise, and
finite waveform-generator bandwidth may alter the ranking of candidate pulses.
The optimized controls should therefore be filtered through measured transfer
functions and recalibrated before experimental use.

The interferometer simulation includes thermal Doppler shifts, transverse
intensity inhomogeneity, and an independently specified systematic detuning.
It does not include atom-atom interactions, wave-front aberrations, detection
noise, or vibration noise.  Consequently, the calculated contrast is a
controlled comparison of pulse-induced effects rather than a prediction of
the absolute sensitivity of a complete gravimeter.  This boundary is important:
an increase in simulated contrast can improve phase readout, but it does not by
itself establish an equivalent improvement in field sensitivity.

\section{Conclusion}
\label{sec:conclusion}

This study built a reproducible optimal-control benchmark for robust Raman mirror pulses under a common peak-amplitude constraint. Projected Krotov produced a slightly larger high-fidelity robustness region than GRAPE, whereas GRAPE achieved a marginally lower terminal error and the largest simulated interferometer contrast. CRAB seed-to-seed variation was retained and reported, but its limited-budget results were weaker in this protocol. The central conclusion is that robust pulse design should be posed as a reproducible multi-metric optimal-control problem, not as a claim of universal superiority for a single optimizer.

Future work should validate the optimized mirror pulses with measured optical
transfer functions and extend the joint optimization to the two beam-splitter
pulses.  An open-system three-level treatment and experimentally measured
noise spectra would provide the next step toward apparatus-specific controls.

\section*{Data and Code Availability}

The reproducible Python pipeline records the complete configuration, optimized
controls, random seeds, software versions, source data, and figure outputs for
each run.  These materials will be deposited in a public repository upon
publication.

%---------------------------------------------------------------------------------------------

% \begin{acknowledgments}
% We wish to acknowledge the support of the author community in using
% REV\TeX{}, offering suggestions and encouragement, testing new versions,
% \dots.
% \end{acknowledgments}

\appendix

\bibliographystyle{apsrev4-2} %格式
\bibliography{apssamp}% Produces the bibliography via BibTeX.

@PREAMBLE{
 "\providecommand{\noopsort}[1]{}" 
 # "\providecommand{\singleletter}[1]{#1}%" 
}

@article{Flechtner_2021,
  title         = {Satellite Gravimetry: A Review of Its Realization},
  author        = {Flechtner, Frank and Reigber, Christoph and Rummel, Reiner and Balmino, Georges},
  year          = {2021},
  month         = sep,
  journal       = {Surveys in Geophysics},
  publisher     = {Springer Science and Business Media LLC},
  volume        = {42},
  number        = {5},
  pages         = {1029–1074},
  doi           = {10.1007/s10712-021-09658-0},
  issn          = {1573-0956},
  url           = {http://dx.doi.org/10.1007/s10712-021-09658-0}
}

@article{Jensen_2025,
  title         = {Airborne gravimetry with quantum technology: observations from Iceland and Greenland},
  author        = {Jensen, Tim Enzlberger and Dale, Bjørnar and Stokholm, Andreas and Forsberg, René and Bresson, Alexandre and Zahzam, Nassim and Bonnin, Alexis and Bidel, Yannick},
  year          = {2025},
  month         = apr,
  journal       = {Earth System Science Data},
  publisher     = {Copernicus GmbH},
  volume        = {17},
  number        = {4},
  pages         = {1667–1684},
  doi           = {10.5194/essd-17-1667-2025},
  issn          = {1866-3516},
  url           = {http://dx.doi.org/10.5194/essd-17-1667-2025}
}

@article{Rosi_2014,
  title         = {Precision measurement of the Newtonian gravitational constant using cold atoms},
  author        = {Rosi, G. and Sorrentino, F. and Cacciapuoti, L. and Prevedelli, M. and Tino, G. M.},
  year          = {2014},
  month         = jun,
  journal       = {Nature},
  publisher     = {Springer Science and Business Media LLC},
  volume        = {510},
  number        = {7506},
  pages         = {518–521},
  doi           = {10.1038/nature13433},
  issn          = {1476-4687},
  url           = {http://dx.doi.org/10.1038/nature13433}
}

@article{Lamporesi_2008,
  title         = {Determination of the Newtonian Gravitational Constant Using Atom Interferometry},
  author        = {Lamporesi, G. and Bertoldi, A. and Cacciapuoti, L. and Prevedelli, M. and Tino, G. M.},
  year          = {2008},
  month         = feb,
  journal       = {Physical Review Letters},
  publisher     = {American Physical Society (APS)},
  volume        = {100},
  number        = {5},
  doi           = {10.1103/physrevlett.100.050801},
  issn          = {1079-7114},
  url           = {http://dx.doi.org/10.1103/PhysRevLett.100.050801}
}

@article{Canuel_2006,
  title         = {Six-Axis Inertial Sensor Using Cold-Atom Interferometry},
  author        = {Canuel, B. and Leduc, F. and Holleville, D. and Gauguet, A. and Fils, J. and Virdis, A. and Clairon, A. and Dimarcq, N. and Bordé, Ch. J. and Landragin, A. and Bouyer, P.},
  year          = {2006},
  month         = jul,
  journal       = {Physical Review Letters},
  publisher     = {American Physical Society (APS)},
  volume        = {97},
  number        = {1},
  doi           = {10.1103/physrevlett.97.010402},
  issn          = {1079-7114},
  url           = {http://dx.doi.org/10.1103/PhysRevLett.97.010402}
}

@article{Cheiney_2018,
  title         = {Navigation-Compatible Hybrid Quantum Accelerometer Using a Kalman Filter},
  author        = {Cheiney, Pierrick and Fouché, Lauriane and Templier, Simon and Napolitano, Fabien and Battelier, Baptiste and Bouyer, Philippe and Barrett, Brynle},
  year          = {2018},
  month         = sep,
  journal       = {Physical Review Applied},
  publisher     = {American Physical Society (APS)},
  volume        = {10},
  number        = {3},
  doi           = {10.1103/physrevapplied.10.034030},
  issn          = {2331-7019},
  url           = {http://dx.doi.org/10.1103/PhysRevApplied.10.034030}
}

@article{Dickerson_2013,
  title         = {Multiaxis Inertial Sensing with Long-Time Point Source Atom Interferometry},
  author        = {Dickerson, Susannah M. and Hogan, Jason M. and Sugarbaker, Alex and Johnson, David M. S. and Kasevich, Mark A.},
  year          = {2013},
  month         = aug,
  journal       = {Physical Review Letters},
  publisher     = {American Physical Society (APS)},
  volume        = {111},
  number        = {8},
  doi           = {10.1103/physrevlett.111.083001},
  issn          = {1079-7114},
  url           = {http://dx.doi.org/10.1103/PhysRevLett.111.083001}
}

@article{Kasevich_1991,
  title         = {Atomic interferometry using stimulated Raman transitions},
  author        = {Kasevich, Mark and Chu, Steven},
  year          = {1991},
  month         = jul,
  journal       = {Physical Review Letters},
  publisher     = {American Physical Society (APS)},
  volume        = {67},
  number        = {2},
  pages         = {181–184},
  doi           = {10.1103/physrevlett.67.181},
  issn          = {0031-9007},
  url           = {http://dx.doi.org/10.1103/PhysRevLett.67.181}
}

@article{Ekstrom_1995,
  title         = {Measurement of the electric polarizability of sodium with an atom interferometer},
  author        = {Ekstrom, Christopher R. and Schmiedmayer, Jörg and Chapman, Michael S. and Hammond, Troy D. and Pritchard, David E.},
  year          = {1995},
  month         = may,
  journal       = {Physical Review A},
  publisher     = {American Physical Society (APS)},
  volume        = {51},
  number        = {5},
  pages         = {3883–3888},
  doi           = {10.1103/physreva.51.3883},
  issn          = {1094-1622},
  url           = {http://dx.doi.org/10.1103/PhysRevA.51.3883}
}

@article{Bouchendira_2011,
  title         = {New Determination of the Fine Structure Constant and Test of the Quantum Electrodynamics},
  author        = {Bouchendira, Rym and Cladé, Pierre and Guellati-Khélifa, Saïda and Nez, François and Biraben, François},
  year          = {2011},
  month         = feb,
  journal       = {Physical Review Letters},
  publisher     = {American Physical Society (APS)},
  volume        = {106},
  number        = {8},
  doi           = {10.1103/physrevlett.106.080801},
  issn          = {1079-7114},
  url           = {http://dx.doi.org/10.1103/PhysRevLett.106.080801}
}

@article{Bonnin_2015,
  title         = {Characterization of a simultaneous dual-species atom interferometer for a quantum test of the weak equivalence principle},
  author        = {Bonnin, A. and Zahzam, N. and Bidel, Y. and Bresson, A.},
  year          = {2015},
  month         = aug,
  journal       = {Physical Review A},
  publisher     = {American Physical Society (APS)},
  volume        = {92},
  number        = {2},
  doi           = {10.1103/physreva.92.023626},
  issn          = {1094-1622},
  url           = {http://dx.doi.org/10.1103/PhysRevA.92.023626}
}

@article{Asenbaum_2020,
  title         = {Atom-Interferometric Test of the Equivalence Principle at the 10-12 Level},
  author        = {Asenbaum, Peter and Overstreet, Chris and Kim, Minjeong and Curti, Joseph and Kasevich, Mark A.},
  year          = {2020},
  month         = nov,
  journal       = {Physical Review Letters},
  publisher     = {American Physical Society (APS)},
  volume        = {125},
  number        = {19},
  doi           = {10.1103/physrevlett.125.191101},
  issn          = {1079-7114},
  url           = {http://dx.doi.org/10.1103/PhysRevLett.125.191101}
}

@article{Wu_2019,
  title         = {Gravity surveys using a mobile atom interferometer},
  author        = {Wu, Xuejian and Pagel, Zachary and Malek, Bola S. and Nguyen, Timothy H. and Zi, Fei and Scheirer, Daniel S. and Müller, Holger},
  year          = {2019},
  month         = sep,
  journal       = {Science Advances},
  publisher     = {American Association for the Advancement of Science (AAAS)},
  volume        = {5},
  number        = {9},
  doi           = {10.1126/sciadv.aax0800},
  issn          = {2375-2548},
  url           = {http://dx.doi.org/10.1126/sciadv.aax0800}
}

@article{Szigeti_2012,
  title         = {Why momentum width matters for atom interferometry with Bragg pulses},
  author        = {Szigeti, S S and Debs, J E and Hope, J J and Robins, N P and Close, J D},
  year          = {2012},
  month         = feb,
  journal       = {New Journal of Physics},
  publisher     = {IOP Publishing},
  volume        = {14},
  number        = {2},
  pages         = {023009},
  doi           = {10.1088/1367-2630/14/2/023009},
  issn          = {1367-2630},
  url           = {http://dx.doi.org/10.1088/1367-2630/14/2/023009}
}

@article{Le_Gou_t_2008,
  title         = {Limits to the sensitivity of a low noise compact atomic gravimeter},
  author        = {Le Gouët, J. and Mehlstäubler, T.E. and Kim, J. and Merlet, S. and Clairon, A. and Landragin, A. and Pereira Dos Santos, F.},
  year          = {2008},
  month         = jun,
  journal       = {Applied Physics B},
  publisher     = {Springer Science and Business Media LLC},
  volume        = {92},
  number        = {2},
  pages         = {133–144},
  doi           = {10.1007/s00340-008-3088-1},
  issn          = {1432-0649},
  url           = {http://dx.doi.org/10.1007/s00340-008-3088-1}
}

@article{J_ger_2014,
  title         = {Optimal quantum control of Bose-Einstein condensates in magnetic microtraps: Comparison of gradient-ascent-pulse-engineering and Krotov optimization schemes},
  author        = {Jäger, Georg and Reich, Daniel M. and Goerz, Michael H. and Koch, Christiane P. and Hohenester, Ulrich},
  year          = {2014},
  month         = sep,
  journal       = {Physical Review A},
  publisher     = {American Physical Society (APS)},
  volume        = {90},
  number        = {3},
  doi           = {10.1103/physreva.90.033628},
  issn          = {1094-1622},
  url           = {http://dx.doi.org/10.1103/PhysRevA.90.033628}
}

@article{Daems_2013,
  title         = {Robust Quantum Control by a Single-Shot Shaped Pulse},
  author        = {Daems, D. and Ruschhaupt, A. and Sugny, D. and Guérin, S.},
  year          = {2013},
  month         = jul,
  journal       = {Physical Review Letters},
  publisher     = {American Physical Society (APS)},
  volume        = {111},
  number        = {5},
  doi           = {10.1103/physrevlett.111.050404},
  issn          = {1079-7114},
  url           = {http://dx.doi.org/10.1103/PhysRevLett.111.050404}
}

@article{van_Frank_2014,
  title         = {Interferometry with non-classical motional states of a Bose–Einstein condensate},
  author        = {van Frank, S. and Negretti, A. and Berrada, T. and Bücker, R. and Montangero, S. and Schaff, J.-F. and Schumm, T. and Calarco, T. and Schmiedmayer, J.},
  year          = {2014},
  month         = may,
  journal       = {Nature Communications},
  publisher     = {Springer Science and Business Media LLC},
  volume        = {5},
  number        = {1},
  doi           = {10.1038/ncomms5009},
  issn          = {2041-1723},
  url           = {http://dx.doi.org/10.1038/ncomms5009}
}

@article{Reich_2012,
  title         = {Monotonically convergent optimization in quantum control using Krotov’s method},
  author        = {Reich, Daniel M. and Ndong, Mamadou and Koch, Christiane P.},
  year          = {2012},
  month         = mar,
  journal       = {The Journal of Chemical Physics},
  publisher     = {AIP Publishing},
  volume        = {136},
  number        = {10},
  doi           = {10.1063/1.3691827},
  issn          = {1089-7690},
  url           = {http://dx.doi.org/10.1063/1.3691827}
}

@article{Glaser_2015,
  title         = {Training Schrödinger’s cat: quantum optimal control: Strategic report on current status, visions and goals for research in Europe},
  author        = {Glaser, Steffen J. and Boscain, Ugo and Calarco, Tommaso and Koch, Christiane P. and Köckenberger, Walter and Kosloff, Ronnie and Kuprov, Ilya and Luy, Burkhard and Schirmer, Sophie and Schulte-Herbrüggen, Thomas and Sugny, Dominique and Wilhelm, Frank K.},
  year          = {2015},
  month         = dec,
  journal       = {The European Physical Journal D},
  publisher     = {Springer Science and Business Media LLC},
  volume        = {69},
  number        = {12},
  doi           = {10.1140/epjd/e2015-60464-1},
  issn          = {1434-6079},
  url           = {http://dx.doi.org/10.1140/epjd/e2015-60464-1}
}

@article{Saywell_2023,
  title         = {Enhancing the sensitivity of atom-interferometric inertial sensors using robust control},
  author        = {Saywell, Jack C. and Carey, Max S. and Light, Philip S. and Szigeti, Stuart S. and Milne, Alistair R. and Gill, Karandeep S. and Goh, Matthew L. and Perunicic, Viktor S. and Wilson, Nathanial M. and Macrae, Calum D. and Rischka, Alexander and Everitt, Patrick J. and Robins, Nicholas P. and Anderson, Russell P. and Hush, Michael R. and Biercuk, Michael J.},
  year          = {2023},
  month         = nov,
  journal       = {Nature Communications},
  publisher     = {Springer Science and Business Media LLC},
  volume        = {14},
  number        = {1},
  doi           = {10.1038/s41467-023-43374-0},
  issn          = {2041-1723},
  url           = {http://dx.doi.org/10.1038/s41467-023-43374-0}
}

@book{Shore_2011,
  title         = {Manipulating Quantum Structures Using Laser Pulses},
  author        = {Shore, Bruce W.},
  year          = {2011},
  month         = sep,
  publisher     = {Cambridge University Press},
  doi           = {10.1017/cbo9780511675713},
  isbn          = {9780511675713},
  url           = {http://dx.doi.org/10.1017/CBO9780511675713}
}

@book{Paul_1997,
  title         = {Atom Interferometry},
  author        = {Paul R. Berman},
  year          = {1997},
  publisher     = {Cambridge University Press},
  doi           = {10.1016/B978-0-12-092460-8.X5000-0},
  isbn          = {978-0-12-092460-8},
  url           = {https://doi.org/10.1016/B978-0-12-092460-8.X5000-0}
}

@article{Saywell_2018,
  title         = {Optimal control of mirror pulses for cold-atom interferometry},
  author        = {Saywell, Jack C. and Kuprov, Ilya and Goodwin, David and Carey, Max and Freegarde, Tim},
  year          = {2018},
  month         = aug,
  journal       = {Physical Review A},
  publisher     = {American Physical Society (APS)},
  volume        = {98},
  number        = {2},
  doi           = {10.1103/physreva.98.023625},
  issn          = {2469-9934},
  url           = {http://dx.doi.org/10.1103/PhysRevA.98.023625}
}

@article{Dunning_2014,
  title         = {Composite pulses for interferometry in a thermal cold atom cloud},
  author        = {Dunning, Alexander and Gregory, Rachel and Bateman, James and Cooper, Nathan and Himsworth, Matthew and Jones, Jonathan A. and Freegarde, Tim},
  year          = {2014},
  month         = sep,
  journal       = {Physical Review A},
  publisher     = {American Physical Society (APS)},
  volume        = {90},
  number        = {3},
  doi           = {10.1103/physreva.90.033608},
  issn          = {1094-1622},
  url           = {http://dx.doi.org/10.1103/PhysRevA.90.033608}
}

@article{L_pez_Monjaraz_2024,
  title         = {Impact of the pi-pulse shape on the contrast of thermal Atom Interferometers},
  author        = {López-Monjaraz, Cristian and Peña Vega, Hellmunt and Jiménez-García, Karina and López Romero, José M and Corzo, Neil V},
  year          = {2024},
  month         = nov,
  journal       = {Physica Scripta},
  publisher     = {IOP Publishing},
  volume        = {99},
  number        = {12},
  pages         = {125414},
  doi           = {10.1088/1402-4896/ad92b2},
  issn          = {1402-4896},
  url           = {http://dx.doi.org/10.1088/1402-4896/ad92b2}
}

@article{Magnus_1954,
  title         = {On the exponential solution of differential equations for a linear operator},
  author        = {Magnus, Wilhelm},
  year          = {1954},
  month         = nov,
  journal       = {Communications on Pure and Applied Mathematics},
  publisher     = {Wiley},
  volume        = {7},
  number        = {4},
  pages         = {649–673},
  doi           = {10.1002/cpa.3160070404},
  issn          = {1097-0312},
  url           = {http://dx.doi.org/10.1002/cpa.3160070404}
}

@article{Miao_2000,
  title         = {An explicit criterion for existence of the Magnus solution for a coupled spin system under a time-dependent radiofrequency pulse},
  author        = {Miao, Xijia},
  year          = {2000},
  month         = jul,
  journal       = {Physics Letters A},
  publisher     = {Elsevier BV},
  volume        = {271},
  number        = {4},
  pages         = {296–302},
  doi           = {10.1016/s0375-9601(00)00373-x},
  issn          = {0375-9601},
  url           = {http://dx.doi.org/10.1016/S0375-9601(00)00373-X}
}

@article{Luo_2016,
  title         = {Contrast enhancement via shaped Raman pulses for thermal cold atom cloud interferometry},
  author        = {Luo, Yukun and Yan, Shuhua and Hu, Qingqing and Jia, Aiai and Wei, Chunhua and Yang, Jun},
  year          = {2016},
  month         = dec,
  journal       = {The European Physical Journal D},
  publisher     = {Springer Science and Business Media LLC},
  volume        = {70},
  number        = {12},
  doi           = {10.1140/epjd/e2016-70428-6},
  issn          = {1434-6079},
  url           = {http://dx.doi.org/10.1140/epjd/e2016-70428-6}
}

@article{Brzozowski_2002,
  title         = {Time-of-flight measurement of the temperature of cold atoms for short trap-probe beam distances},
  author        = {Brzozowski, Tomasz M and Maczynska, Maria and Zawada, Michal and Zachorowski, Jerzy and Gawlik, Wojciech},
  year          = {2002},
  month         = jan,
  journal       = {Journal of Optics B: Quantum and Semiclassical Optics},
  publisher     = {IOP Publishing},
  volume        = {4},
  number        = {1},
  pages         = {62–66},
  doi           = {10.1088/1464-4266/4/1/310},
  issn          = {1741-3575},
  url           = {http://dx.doi.org/10.1088/1464-4266/4/1/310}
}

@article{Stoner_2011,
  title         = {Analytical framework for dynamic light pulse atom interferometry at short interrogation times},
  author        = {Stoner, Richard and Butts, David and Kinast, Joseph and Timmons, Brian},
  year          = {2011},
  month         = sep,
  journal       = {Journal of the Optical Society of America B},
  publisher     = {Optica Publishing Group},
  volume        = {28},
  number        = {10},
  pages         = {2418},
  doi           = {10.1364/josab.28.002418},
  issn          = {1520-8540},
  url           = {http://dx.doi.org/10.1364/JOSAB.28.002418}
}

@article{Butts_2011,
  title         = {Light pulse atom interferometry at short interrogation times},
  author        = {Butts, David L. and Kinast, Joseph M. and Timmons, Brian P. and Stoner, Richard E.},
  year          = {2011},
  month         = feb,
  journal       = {Journal of the Optical Society of America B},
  publisher     = {Optica Publishing Group},
  volume        = {28},
  number        = {3},
  pages         = {416},
  doi           = {10.1364/josab.28.000416},
  issn          = {1520-8540},
  url           = {http://dx.doi.org/10.1364/JOSAB.28.000416}
}

@article{Wang_2024,
  title         = {Amplitude-modulated mirror pulses for improved cold-atom gravimetry},
  author        = {Wang, Yida and Cheng, Jinshuo and Liu, Yujuan and Lin, Tingting},
  year          = {2024},
  month         = may,
  journal       = {Physical Review A},
  publisher     = {American Physical Society (APS)},
  volume        = {109},
  number        = {5},
  doi           = {10.1103/physreva.109.053501},
  issn          = {2469-9934},
  url           = {http://dx.doi.org/10.1103/PhysRevA.109.053501}
}

@article{Palao_2003,
  title         = {Optimal control theory for unitary transformations},
  author        = {Palao, José P. and Kosloff, Ronnie},
  year          = {2003},
  month         = dec,
  journal       = {Physical Review A},
  publisher     = {American Physical Society (APS)},
  volume        = {68},
  number        = {6},
  doi           = {10.1103/physreva.68.062308},
  issn          = {1094-1622},
  url           = {http://dx.doi.org/10.1103/PhysRevA.68.062308}
}

@article{Khaneja_2005,
  author  = {Khaneja, Navin and Reiss, Timo and Kehlet, Cindie and Schulte-Herbrueggen, Thomas and Glaser, Steffen J.},
  title   = {Optimal control of coupled spin dynamics: Design of {NMR} pulse sequences by gradient ascent algorithms},
  journal = {Journal of Magnetic Resonance},
  volume  = {172},
  number  = {2},
  pages   = {296--305},
  year    = {2005},
  doi     = {10.1016/j.jmr.2004.11.004}
}

@article{Caneva_2011,
  author  = {Caneva, Tommaso and Calarco, Tommaso and Montangero, Simone},
  title   = {Chopped random-basis quantum optimization},
  journal = {Physical Review A},
  volume  = {84},
  pages   = {022326},
  year    = {2011},
  doi     = {10.1103/PhysRevA.84.022326}
}

@article{Johansson_2012,
  author  = {Johansson, J. R. and Nation, P. D. and Nori, Franco},
  title   = {{QuTiP}: An open-source {Python} framework for the dynamics of open quantum systems},
  journal = {Computer Physics Communications},
  volume  = {183},
  number  = {8},
  pages   = {1760--1772},
  year    = {2012},
  doi     = {10.1016/j.cpc.2012.02.021}
}
% \bibliography{apsrev4-2}
\end{document}